\documentstyle[12pt]{article}

            \def\beq{\begin{equation}}
\def\eeq#1{\label{#1}\end{equation}}            
    \def\dfrac#1#2{{\displaystyle\frac{#1}{#2}}}

\begin{document}

\title{The Bright Side of Dark Matter\thanks{received an ``honorable mention" in the 1999 Gravity Research Foundation Essay Competition.}}   
\author{Ariel Edery\thanks{E-mail: edery@hep.physics.mcgill.ca}\\Department of Physics, McGill University\\3600 University 
Street\\Montreal, PQ, Canada, H3A 2T8}
\date{}
\maketitle
\begin{abstract} 
We show that it is not possible in the absence of dark matter to construct a four-dimensional metric
that explains galactic observations. In particular, by working with an effective potential it is shown that a metric which is 
constructed to fit flat rotation curves in spiral galaxies leads to the wrong sign for the bending of light i.e. repulsion instead of attraction. 
Hence, without dark matter the motion of particles on galactic scales cannot be explained in terms of geodesic motion on a 
four-dimensional metric. This reveals a new bright side to dark matter: it is 
indispensable if we wish to retain the cherished equivalence principle.

\end{abstract}

%\footnotetext[1]{ Department of Physics, McGill
%University, 3600 University St., Montr\'eal, Qu\'ebec, 
%H3A-2T8 Canada. E-mail: edery@hep.physics.mcgill.ca}

%\title{The bright side of dark matter}
%\author{A. Edery${}^{1}$ }

%\date{ }

%\maketitle

It has been known for over sixty years, since the work of Zwicky \cite{Zwicky} in the early 1930s, that there are significant discrepancies 
between the luminous and dynamical 
mass in large astronomical systems such as galaxies and clusters of galaxies. The luminous mass in galaxies is considerably less than
the dynamical mass inferred from applying Newtonian gravity to the motion of gas and stars orbiting the galaxies. In clusters of 
galaxies the luminous mass is again considerably less than the dynamical mass inferred from applying the virial theorem to the motion of the galaxies (keeping in mind that 
correctly applying the virial theorem depends on whether the clusters are in equilibrium). Gravitational lensing also reveals a mass 
discrepancy: the dynamical mass inferred from applying General Relativity to the bending of light in clusters is 
considerably greater than the luminous mass. To date, the most 
detailed evidence for the mass discrepancy in galaxies is
the extended rotation curves in spiral galaxies determined from the observed 21 cm line emission of neutral hydrogen; outside the
optical disc the rotation curves remain flat instead of falling off in a Keplerian fashion. There seems to be only two possible 
explanations for the mass discrepancy in
astronomical systems. Either there is large amounts of non-luminous matter (or dark matter) that clusters on galactic scales or the
Newtonian inverse-square law used to infer the dynamical mass breaks down on galactic scales (which implies General Relativity breaks 
down on those 
scales too). At the present time the dark matter paradigm is by far the most favored option. Nonetheless, the direct experimental 
detection of either baryonic or nonbaryonic dark matter in the amounts inferred from observations is presently lacking. The nature of 
the dark matter is still unknown. A few authors have therefore considered 
alternative gravity as a possible solution to the mass discrepancy; the best known of these is Milgrom's Modified Newtonian 
Dynamics (MOND) \cite{Milgrom}. In MOND, Newtonian
dynamics are modified at low accelerations typical of orbits on galactic scales. It has been reasonably successful at fitting galactic
rotation curves using only one extra parameter. The main drawback with MOND is that it is a non-relativistic theory and as such cannot 
make any predictions on cosmology, the deflection of light, etc. Attempts at constructing a relativistic theory based on MOND have not been successful \cite{Bekenstein}. Other 
proposed alternatives include Bekenstein and Sanders' scalar-tensor theory \cite{BekSand}. However, it was found that the scalar field contributed a negative deflection of light 
which reduced the overall deflection instead of increasing it. To date, no satisfactory alternative to General Relativity has been 
found that can explain the galactic observations (in our context ``galactic observations" means the galactic rotation curves and the 
observed bending of light). After over sixty years the problem of the mass discrepancy in astronomical systems is still with us and stands as one of the 
great unsolved problems in astrophysics.

In this letter we shed light on the problem by showing that there is a serious price to pay for choosing alternative gravity over 
dark matter: the equivalence principle must be violated.
We arrive at this conclusion by first addressing the following question: without introducing dark matter can one construct a metric which is 
a modification of the 
Schwarzschild metric and explain galactic phenomenology? We show that such a metric does not exist. 
By analyzing the motion of particles with an 
effective potential it is shown that a metric constructed to explain flat 
rotation curves in spiral galaxies leads to light repulsion! 
Hence, without dark matter galactic observations cannot be explained in terms of geodesic 
motion on a four-dimensional 
metric. Simply put, dropping dark matter 
for alternative gravity implies doing away with the equivalence principle.

We now prove our conjecture that without dark matter a metric which explains flat rotation curves leads to the wrong sign for the
bending of light. We assume that the total mass $M$ of a spiral galaxy is luminous and is found inside the optical disc (we
neglect the mass of the gaseous component which usually makes a negligible contribution compared to the total luminous mass). Therefore,   
outside the optical disc the mass within a sphere of radius $r$ is constant and equal to $M$. Under the 
usual inverse $r$ potential associated with the Schwarzschild metric the velocity of the gas orbiting the outskirts of the galaxy would 
not remain 
constant but  
decrease as a function of $r$. Therefore, we expect on galactic scales a potential that actually increases with distance in order to 
explain the flat rotation curves.  

Consider a static
spherically symmetric metric $ds^2 = B(r) dt^2 - A(r) dr^2 - r^2 d\Omega^2$ with $B(r) \equiv 1+2\phi(r)$ (we will specify the function $A(r)$ later). The velocity $v$ of gas in circular
orbits around a galaxy is independent of the function $A(r)$ and is given by 
$v^2 = r\, \phi'(r)$ \cite{Weinberg}. Since the 
velocity of the gas is constant (flat) on the outskirts of spiral galaxies, the function $\phi(r)$ on galactic scales is given by 
$C \,ln(r/b)$ where $C$ and $b$ are positive parameters adjusted for each galaxy.  A logarithmic function on galactic 
scales is therefore best suited to explain the flat rotation curves in the
absence of dark matter. The dimensionless constant $C$ is of order 
$v^{2}/c^{2}$ and since spiral galaxies have flat rotation velocities $v$ in the range 60 to 300 km/s \cite{Sanders}, the value of $C$ roughly ranges from $10^{-8}$ to $10^{-6}$. Actually the constant $C$ is constrained by the 
Tully-Fisher relation to be proportional to the square root of the mass $M$ of the galaxy \cite{Milbek} and this has been shown to fit rotation curves 
quite well \cite{Sanders}). 

Of course, we need to recover the results of the Schwarzschild metric on solar system scales (such as the deflection of light from the sun, the precession of the perihelia of Mercury, etc.). We therefore let
$B(r)= 1-2GM/r +2C\,ln(r/b)$ (the constant $C$ is small enough so that the logarithmic term
is negligible on solar system scales and similarly the $GM$ term is negligible when $r$ is large i.e. on galactic scales). 
We also require to a very high degree the same relation between $A(r)$ and 
$B(r)$ as in the Schwarzschild metric i.e. $A(r)=B^{-1}(r)$. 
Deviations from this relation spoil the classical solar system tests. The important point is that a 
metric which reduces to the Schwarzschild metric when $r$ is small but which does not have the relation $A(r)=B^{-1}(r)$ cannot in general 
reproduce the solar system results. 
To see this consider a metric with $B(r)=1-2GM/r + 2f(r)$ and $A(r)= 1/(1-2GM/r + g(r))$ where $f(r)$ and $g(r)$ are functions that are
small compared to $GM/r$ on solar system scales. Moreover, to reproduce galactic rotation curves the function $f(r)$ will be an increasing function of $r$ and will 
approach infinity asymptotically (this is certainly the case for the logarithmic function we obtained but we wish here to be more general and not specify $f(r)$ exactly). The deflection of
light with these two functions can be calculated using the standard integral-angle formula \cite{Weinberg} and yields:  
\begin{eqnarray}
\Delta \varphi = &2& \int_{0}^{\pi/2} \sqrt{A(r)\,B(r)}   \\ \nonumber
&\,&\left(1 + \dfrac{GM}{r_{0}}\dfrac{(1-\sin^{3}\theta)}{\cos^{2}\theta}
- f(r_{0}) + (f(r)-f(r_{0})) \,\tan^{2}\theta \right) d\theta  - \pi 
\end{eqnarray}
where $r_{0}$ is the point of closest approach, $r=r_{0}/\sin\theta\,$ and terms inside the brackets must be much smaller than the constant one 
to obtain any realistic deflection (of order arc seconds). The integration from $0$ to $\pi/2$ represents light moving from $r_{0}$ to infinity. Note that the $GM$ term and the $f(r)$ terms are 
separated inside the brackets i.e. there are no cross terms. If the product  
$A(r)\,B(r)$ is equal to one then the straight line angle $\pi$ is cancelled out exactly and one obtains
the deflection $4GM/r_{0}$ as in the Schwarzschild case plus a modification proportional to $f(r_{0})$ (which by definition plays a significant role only when $r_{0}$ is on 
galactic scales and hence does not spoil the ``Schwarzschild" result). However, if $A(r)\,B(r) \ne 1$ this changes the situation in a significant way.  First, if $g(r)$ does not approach infinity
asymptotically the deflection is infinite. Secondly, even if $g(r)$ approaches infinity asymptotically and the deflection is finite it will not reproduce the solar system results if
$A(r)\,B(r)$ is not equal to one. The straight-line angle $\pi$ will in general no longer cancel out as before and a constant deflection term will appear (which has not been observed on solar scales). 
Moreover, the result $4GM/r_{0}$ will no longer be reproduceable and cross terms involving combinations of $GM/r_{0}$, $f(r_{0})$ and $g(r_{0})$ will now appear. Since the solar deflection result 
$4GM/r_{0}$ has been confirmed to an accuracy of 1\% using radio-interferometric methods \cite{Fomalont} this leaves very little room for any deviation from a $A(r)=B^{-1}(r)$ relation. 
  
We now show that the deflection of light on galactic scales is negative. One way to proceed is to calculate the integral 
for the deflection angle and show that the extra term is negative (this has already been done for the case where $\phi(r)$ is a linear function of $r$ \cite{Edery,Walker}). However, such a calculation is 
not necessary nor illuminating. It proves much more instructive to work with an effective potential. As in the analysis of the Schwarzschild metric   
one can write the geodesic equations of motion as ``Newtonian" equations of motion with an 
effective potential $V(r)$ \cite{Edery, Wald}. The derivative of the potential can be
interpreted as the ``radial force"  acting on the 
particles and therefore its sign reveals whether a particle is attracted or 
repelled by the source. Armed with a potential we will have no need to calculate the full deflection angle and this will  enable us to carry out a clear and  
general analysis. The effective potential for a metric with $B(r)=1+2\phi(r)$ and $A(r)=B^{-1}(r)$ is given by \cite{Edery}
\beq
V(r)= \phi(r)\left(\dfrac{J^{2}}{r^{2}} + E \right)
\eeq{potential}
where $J$ and $E$ are constants of the motion. The above potential is a velocity-dependent potential. For particles moving 
non-relativistically (NR for short), such as gas in circular orbits around a spiral galaxy, one has the condition $E\approx 1$ and 
$J^{2}/r^{2}<<1$ yielding the potential $V_{NR}(r) \approx \phi(r)$. For light, $E$ is zero and the potential is 
$V_{light}(r) = \phi(r) (J^{2}/r^{2})$.  
We now substitute the logarithmic function $\phi(r)=C\,ln(r/b)$ into these two potentials. For the case of light, we can relate the constant $b$ to the point of 
closest approach $r_{0}$ by using the fact that the derivative of $V_{light}(r)$ is a maximum at $r_{0}$. This leads to $b=r_{0}e^{-5/6}$. The derivative of $V_{light}(r)$ 
can now be readily calculated and is found to be negative corresponding to repulsion. Note that $V_{light}(r)$ decreases with $r$. In contrast, the NR potential, which is equal to the logarithmic function, increases with $r$ and its derivative is 
positive corresponding to attraction.  
We therefore see that the logarithmic function required to explain flat rotation curves in spiral galaxies
leads to light repulsion. It is worthwhile to note that the deflection of light is a well defined finite quantity for the given 
logarithmic function $\phi(r)$ 
precisely because the potential for light decreases with $r$. If the potential for light had been an increasing function of $r$ the
deflection of light would yield a nonsensical divergent result i.e. scattering states would simply not exist.
Hence, we see immediately that for functions $\phi(r)$ that increase faster than $r^{2}$ one cannot even talk of a deflection.  
Note that for a function $\phi(r)= -GM/r$ the derivative of both potentials is positive as required on solar system scales. 

It is important to note that a negative deflection is not particular to a logarithmic function and this is where using an effective
potential enables us to make a more general and powerful conclusion. We see immediately that the derivative 
of $V_{light}(r)$ is  
negative for any function $\phi(r)$ that increases 
slower than $r^2$ (which is well within the requirement for fitting galactic rotation curves). Hence, 
if we had allowed deviations from a strict logarithmic function in the fitting of galactic rotation curves the 
result would not change: the deflection of light would still be negative. One thing is clear: an increasing function $\phi(r)$ is 
required to explain the flat rotation curves and this inevitably leads to light repulsion. There is no way around this general and 
striking result. We have shown that, in the absence of 
dark matter, it is not possible to construct a four-dimensional metric that explains galactic phenomenology. Therefore, any 
attempt to explain the mass discrepancy in astronomical systems 
using alternative gravity instead of dark matter comes at the 
price of having to abandon the equivalence principle. Dark matter can now be seen in a new light: it is indispensable if we wish to 
hold on to this pillar of gravitational theory.

\section*{ACKNOWLEDGEMENTS}

I wish to thank Robert Myers for his useful 
comments and McGill University for their financial support.   


\begin{thebibliography}{99}

\bibitem{Zwicky} F. Zwicky, Helv. Phys. Acta, {\bf 6}, 110 (1933).
\bibitem{Milgrom} M. Milgrom, Astrophys. J. {\bf 270}, 365,371,384 (1983).
\bibitem{Bekenstein} J. D. Bekenstein, Phys. Lett. B {\bf 202}, 497 (1988).
\bibitem{BekSand} J. D. Bekenstein and R. H. Sanders, Astrophys. J. {\bf 429}, 480 (1994).
\bibitem{Weinberg} S. Weinberg, {\it Gravitation and Cosmology} (Wiley, New York, 1972).
\bibitem{Milbek} J. D. Bekenstein and M. Milgrom, Astrophys. J. {\bf 286}, 7 (1984).
\bibitem{Sanders} R.H. Sanders, Astrophys. J. {\bf 473}, 117 (1996).
\bibitem{Fomalont} E. B. Fomalont and R. A. Sramek, Phys. Rev. Lett. {\bf 36}, 1475 (1976).
\bibitem{Edery} A. Edery and M. B. Paranjape, Phys. Rev. D {\bf 58} 024011(1998).
\bibitem{Walker} M. A. Walker, Astrophys. J. {\bf 430}, 463 (1994).
\bibitem{Wald} R. Wald, {\it General Relativity} (University of Chicago Press, Chicago, 1984).



         

\end{thebibliography}
\end{document}